\RequirePackage{rotating}
\documentclass[useAMS,usenatbib]{mn2e}

\usepackage{graphics}
\usepackage{graphicx}
\usepackage{amsxtra}
\usepackage{journals}
\usepackage{url}
\usepackage{latexsym}
\usepackage{amssymb}
\usepackage{systeme}
\usepackage{times}
\usepackage{amsmath}
\usepackage{subfigure}
\usepackage{natbib}
\usepackage[colorlinks=true, linkcolor=blue, citecolor=blue, urlcolor=blue]{hyperref}
\usepackage{soul,color}
\usepackage{txfonts}
\usepackage{rotating}
\usepackage{lscape} 
\usepackage{threeparttable} 
\usepackage{mathtools} 
\usepackage{ulem}

\usepackage{color}

\begin{document}
\title
[Reddening and extinction for TGAS stars]
{Verifying reddening and extinction for {\it Gaia} DR1 TGAS main sequence stars}
\author[Gontcharov \& Mosenkov.]
{
George~A.~Gontcharov$^{1,2}$\thanks{E-mail: george.gontcharov@tdt.edu.vn} and Aleksandr~V.~Mosenkov$^{3,4,5}$
\newauthor
\\ \\
$^{1}$Department for Management of Science and Technology Development,
Ton Duc Thang University, Ho Chi Minh City, Vietnam\\
$^{2}$Faculty of Applied Sciences, Ton Duc Thang University, Ho Chi Minh City, Vietnam\\
$^{3}$Sterrenkundig Observatorium, Universiteit Gent, Krijgslaan 281, B-9000 Gent, Belgium\\
$^{4}$St.Petersburg State University, 7/9 Universitetskaya nab., St.Petersburg, 199034 Russia\\
$^{5}$Central Astronomical Observatory, Russian Academy of Sciences, 65/1 Pulkovskoye chaussee, St. Petersburg, 196140 Russia\\
}

\date{\today}

\pagerange{\pageref{firstpage}--\pageref{lastpage}} \pubyear{2017}

\maketitle

\label{firstpage}

\begin{abstract}
We compare eight sources of reddening and extinction estimates for approximately 60,000 {\it Gaia} DR1 Tycho--Gaia Astrometric Solution (TGAS)
main sequence stars younger than 3~Gyr with a relative error of the {\it Gaia} parallax less than 0.1.
For the majority of the stars, the best 2D dust emission-based reddening maps show considerable differences between the reddening to infinity and 
the one calculated to the stellar distance using the barometric law of the dust distribution.
This proves that the majority of the TGAS stars are embedded in to the Galactic dust layer and a proper 3D treatment of the reddening/extinction
is required to reliably calculate 
their de-reddened colors and absolute magnitudes.
The sources with the 3D estimates of reddening are tested in their ability to put the stars among the PARSEC and MIST theoretical isochrones in 
the Hertzsprung--Russell diagram
based on the precise {\it Gaia}, {\it Tycho-2}, 2MASS and {\it WISE} photometry.
Only the reddening/extinction estimates by Arenou et al. (1992), Gontcharov (2012), and Gontcharov (2017), being appropriate for nearby stars within 280~pc,
provide both the minimal number of outliers bluer than any reasonable isochrone and 
the correct number of stars younger than 3~Gyr in agreement with the Besan\c{c}on Galaxy model.
\end{abstract}


\section{Introduction}
\label{sec.intro}
Recently, the positions from the {\it Hipparcos} \citep{hip2} and {\it Tycho-2} \citep{tycho2} catalogues have been used as additional information for a joint 
solution with early data of the {\it Gaia} mission and presented as the {\it Gaia} DR1 Tycho--Gaia Astrometric Solution \citep[TGAS, ][]{gaiaa, gaiab}.  
\citet{bailer3} have calculated the most probable distances for the TGAS stars using their parallaxes $\varpi$ with errors $\sigma(\varpi)$.

The {\it Gaia} DR1 itself, {\it Tycho-2}, the Two Micron All-Sky Survey \citep[2MASS,][]{2mass}, the {\it Wide-field Infrared Survey Explorer} \citep[{\it WISE},][]{wise},
the Sloan Digital Sky Survey \citep[SDSS,][]{Eisenstein2011}, the AAVSO Photometric All Sky Survey \citep[APASS,][]{apass},
and other sky surveys provide multi-color photometry for a large number of the TGAS stars at the median precision level of $0.02$~mag.
An estimate of the reddening and interstellar extinction for the TGAS stars with the same level of accuracy is needed in order to precisely place them
in the Hertzsprung--Russell (HR) diagram and to carry out a further detailed analysis.
On the other hand, the distribution of the stars in the diagram with respect to their theoretical stellar isochrones after  estimating their reddening and
extinction can potentially provide us with information about the accuracy of this estimate, which allow us to verify the used source of the reddening/extinction.

There are several ways to estimate the reddening and extinction of the TGAS stars. 
They differ according to their methods, bands (which were used to create them), angular and spatial resolution, and other properties.
The reddening and/or extinction of a TGAS star can be presented as 
a result of a direct fitting of its photometry to its spectral energy distribution (SED), temperature, metallicity, and other key parameters, or
as a result of some measurements for stars which are closely located to the TGAS star.
The latter can be presented as a tabulated map or a set of formulas (commonly referred to as an analytical model),
both dependent on the Galactic longitude $l$ and latitude $b$ to infinity (two-dimensional, 2D maps and models) or to a distance $R$ 
(three-dimensional, 3D maps and models).
In this paper, such 2D maps,  3D maps, 3D models and direct measurements of reddening and/or extinction are considered together.

The difference between the 2D and 3D reddening/extinction data sources is important only for stars inside the dust layer, as the dust extinction differs significantly within this layer.
However, the properties of this layer, such as its vertical offset with respect to the Sun and its scale height, are still poorly known.
They can be estimated by measuring the individual reddening and extinction for the stars inside the layer.

\citet{gm2017} have demonstrated for the {\it Kepler} field that the fitting of both the giants and main-sequence stars to the theoretical isochrones 
allows them to separate the most reliable estimates of the reddening and extinction.

In this study we intend to test several reddening and extinction data sources in their ability to correctly place the TGAS stars in the HR diagram.
We select the TGAS stars with the most precise data whose diagrammatical positions with respect to some theoretical isochrones are the most sensitive
to reddening and extinction.
Based on the results of this fitting, the best up-to-date reddening and extinction estimates can be assigned for the TGAS stars within the space under consideration.

This paper is organized as follows. In the next section we select a sample of the TGAS stars with the most precise parallaxes,
then we define the space for them and briefly describe the reddening/extinction data sources which should be used in this space.
In Sect.~\ref{results}, we put the selected sample of stars in the HR diagram and discuss their location with respect to the theoretical isochrones.
We summarize our main findings and conclusions in Sect.~\ref{conclusions}.

\section{Sample and reddening/extinction data sources}
\label{maps}

In our study we must use stars with the most precise TGAS parallaxes.
Therefore we selected 241,681 TGAS stars with a relative error of the parallax $\sigma(\varpi)/\varpi<0.1$,
adding a systematic uncertainty of 0.3~mas to the declared parallax error following the {\it Gaia} team recommendation \citep{gaiab}.
This translates in to an error of the absolute magnitude of $\sigma(M)<0.2$~mag \citep[][ p. 44]{parenago}.
The mean values are $\sigma(\varpi)/\varpi=0.074$ and $\sigma(M)<0.16$~mag, the median ones are $\sigma(\varpi)/\varpi=0.077$ and 
$\sigma(M)<0.17$~mag.
The limit $\sigma(\varpi)/\varpi<0.1$ corresponds to the distance cut-off of $R<280$~pc due to the strong correlation between $\sigma(\varpi)/\varpi$ and $R$.

\subsection{Extinction law}
\label{law}

Although the data sources we consider in this work contain estimates of the reddening and extinction in different bands,
for consistency, all such estimates in our study, where necessary, are converted into the reddening $E(B-V)$.
The formulas of the conversion are given as:
\begin{equation}
\label{arebv}
A_\mathrm{V}=3.1\,E(B-V)\,,
\end{equation}
\begin{equation}
\label{bvjk}
E(B-V)=1.655\,(J-K_s)\,,
\end{equation}
and
\begin{equation}
\label{akav}
A_\mathrm{K_s}=0.117\,A_\mathrm{V}
\end{equation}
with enough precision following the extinction law (i.e. the dependence of extinction on wavelength) of
\citet{cardelli}, realized, for example, in the PARSEC data base \citep{bressan}
\footnote{\url{http://stev.oapd.inaf.it/cgi-bin/cmd}}.
All the results presented in this study are obtained according to this extinction law.
It differs slightly from other popular laws, such as by \citet{odonnell}, \citet{wd2001}, and \citet{fm2007}.
However, for the bands under consideration (spanning a range of wavelengths from 0.4 to 5.4~$\mu$m) these laws provide similar results:
the difference is always within the precision level of the data used.
The usage of different colors for the HR diagram in this study allows us to test the reliability of the adopted extinction law.

\subsection{The Schlegel, Finkbeiner \& Davis (1998) map}
\label{sfd_map}

Since 1998 the main source (according to the number of citations) for estimating the reddening and extinction of any Galactic
or extragalactic object has been the 2D map from \citet[][hereafter SFD]{sfd}. 
This map was obtained on the basis of the data from the NASA space missions {\it COBE/DIRBE} and {\it IRAS (ISSA)} for the dust emission at
far-infrared (FIR) 100~$\mu$m. This emission was calculated from the observer to infinity (i.e. it is the cumulative emission)
including the whole dust layer along the Galactic midplane.
In fact it is the whole dust \textit{half-}layer to the north or 
to the south due to the position of the Sun nearly in the middle of this layer.
Since the diffuse FIR emission is a direct measure of the column density of the interstellar dust, such
a map can be used as a measure of the extinction to extragalactic objects.
The map is normalized to the $E(B-V)$ reddening (which is cumulative as well) using the color excess 
of 389 elliptical galaxies with $E(B-V)<0.2$~mag assuming the relation~(\ref{arebv}) and the extinction laws from \citet{cardelli} and \citet{odonnell}.

A brief review of the main systematic errors of the SFD map with related references can be found in \citet{astroph}.

For extragalactic objects and Galactic stars outside the dust layer the SFD map can be used as it is.
For stars inside the layer, however, the reddening to infinity must be reduced to their distances.
For this, the spatial distribution of the dust and spatial variations of the extinction law should be taken into account to properly estimate the
reddening/extinction to a given distance.
Since the dust distribution and spatial variations of the extinction law in galaxies (including our own Milky Way) can be quite complex 
\citep[see e.g.,][]{astroph}), the creation of such a 3D map is a major issue.
To date, there are only few robust 3D maps based on the SFD map.

Yet, even the simplest way of reducing any 2D map provides a 3D map with a better estimate of the reddening.
This method is based on the assumption that the density of the dust layer and, accordingly,
the reddening and extinction vary with the vertical coordinate $Z$ (perpendicular to the Galactic midplane), following the barometric law
\citep[][ p. 265]{parenago}:
\begin{equation}
\label{baro}
E(B-V)_\mathrm{R}=E(B-V)\,(1-\mathrm{e}^{-|Z-Z_0|/Z_\mathrm{A}})\,,
\end{equation}
where $E(B-V)_\mathrm{R}$ is the reddening to the distance $R$,
$E(B-V)$ is the reddening to infinity for the same line of sight,
$Z=R\sin(b)$ is the Galactic coordinate of the object along the $Z$-axis in kpc,
$Z_0$ is the vertical offset of the midplane of the dust layer with respect to the Sun in kpc, and
$Z_\mathrm{A}$ is the scale height of the dust layer in kpc.

Various estimates of $Z_0$ and $Z_\mathrm{A}$ found in literature are contradictory \citep{perr} (pp. 470--471, 496--497).
For example, \citet{drispe} accept $Z_\mathrm{A}$ depending on the Galactocentric position and $Z_\mathrm{A}=188$~pc at the Sun.
The Besan\c{c}on Galaxy model \citep[][hereafter BMG]{bmg1} accepts $Z_\mathrm{A}=140$~pc.
On the other hand, based on the SDSS photometry together with the extinction estimate from the SFD map, \citet{juric} concluded that
`practically all the dust is confined to a region within $\approx70$~pc from the Galactic midplane', i.e. $Z_\mathrm{A}<70$~pc.
\citet{av}
estimated $Z_\mathrm{A}^\mathrm{eq}=68\pm6$~pc for the equatorial dust layer and $Z_\mathrm{A}^\mathrm{G}=40\pm10$~pc for
the additional dust layer in the Gould Belt (see the model discussed in Sect.~\ref{Gont_model}). 
These layers intersect near the Sun and locally combine into a joint layer of the total $Z_\mathrm{A}\approx100$~pc.
\citet{kos} estimated $Z_\mathrm{A}=118\pm5$~pc for the general dust layer and $Z_\mathrm{A}=209\pm12$~pc for a
layer of matter producing a diffuse interstellar band, at the wavelength of 0.862~$\mu$m.

The parameter $Z_0$ is rather uncertain as well. 
After a discussion, \citet{drispe} set $Z_0=-14.6$~pc.
For SDSS stars with $100<|Z|<300$~pc, \citet{juric} obtained $Z_0=-25\pm5$~pc.
Based on the spatial density of complete samples of the clump and branch giants from the {\it Tycho-2} catalogue,
\citet{rcg, rgb} obtained a similar estimate for both types of giants 
$Z_0=-13\pm1$~pc.
However, based on the spatial density of a complete sample of OB stars,
\citet{ob} showed that $Z_0$ varies substantially and continuously depending on the age of the reference stars:
from $Z_0=-15$~pc for the stars with the age of $200-400$~Myr, to $Z_0=-30$~pc for the stars younger than 100~Myr.
This difference may be explained by the solar motion with respect to the gas and dust layer producing the stars.

In this study we accept some average estimates of the dust layer parameters: $Z_0=-13$~pc and $Z_\mathrm{A}=100$~pc. 
The SFD map converted into the 3D one by use of the barometric law with these $Z_0$ and $Z_\mathrm{A}$ is hereafter designated as SFD$_R$.

\subsection{The {\it Planck} map}
\label{planck}

Recently, another emission-based 2D reddening map was constructed \citep{planck2014}.
It is based on the emission data from the {\it Planck} mission at 353, 545, and 857~GHz, and the IRAS data at 100~$\mu$m.
This map has higher precision and angular resolution with respect to the SFD map.

To estimate the reddening through the Galactic dust half-layer, the emission was calibrated to $E(B-V)$ for 53,399 quasars deduced from the SDSS data.
Therefore, a benefit of this source compared to the SFD map is that the much larger number of objects was used for its calibration.
It is worth noting, however, that the calibration with the quasars was made only for very low reddening $E(B-V)<0.09$~mag.
For higher reddening, this calibration was tested by the direct comparison between the {\it Planck} $E(B-V)$ calibrated with the quasars 
and $E(B-V)$ derived by \citet{schneider} from the 2MASS photometry for the field of the Taurus and $\rho$ Ophiuchi
molecular clouds. 
It appears that the {\it Planck} reddening has a systematic offset by typically 25 per cent and up to 200~per cent for some lines of sight 
\citep{planck2014}.
The method used by \citet{schneider} has produced reddening maps for some local fields by use of the color excesses of the 2MASS
stars. Yet, this method has not been tested on the full sky.
Therefore it should be used with caution for calibrating emission-based maps.
Thus, such calibration of the {\it Planck} and any other emission-based map is a major issue.

\citet{2015ApJ...798...88M} 
derived the full-sky 6.1-arcmin resolution maps of dust optical depth and temperature by fitting the two-component model of \citet{fds}
to {\it Planck} 217--857~GHz, along with the {\it DIRBE/IRAS} 100~$\mu$m data.
They calibrated the obtained two-component optical depth map to the $E(B-V)$ map (hereafter the PLA map) by use of 
230,000 SDSS stars at $|b|>20^{\circ}$ with both spectroscopy and broadband photometry as proposed by \citet{sf}.
The larger range of the reddening of the stars used ($E(B-V)<0.4$~mag) should provide better quality of the map with respect to the SFD and the 
original {\it Planck} map. 

Similar to the SFD map, in our study the PLA map is converted into the 3D one by use of the barometric law, and it is hereafter referred to as PLA$_R$.

\subsection{The Drimmel, Cabrera-Lavers \& L{\'o}pez-Corredoira (2003) map}
\label{drimmel_map}

\citet[][hereafter DCL]{drimmel} created their 3D Galactic extinction map based on the Galactic dust distribution model
from \citet{drispe}.
This model consists of three structural components: a warped, but otherwise axisymmetric disc with a radial
temperature gradient, spiral arms as mapped by known H~{\sc~ii} regions, and a local Orion-Cygnus arm segment.
Parameters of the model are constrained by the total Galactic emission from {\it DIRBE} observations (for $|b|<30^{\circ}$) or SFD
(for $|b|>30^{\circ}$) FIR maps.
In fact, the {\it DIRBE} and SFD maps differ only by their resolution. Therefore, finally the DCL map is constrained by the SFD
$E(B-V)$ values (to infinity).
To reduce the emission/reddening/extinction between the Sun ($E(B-V)=0$~mag) and the edge of the dust layer
(SFD's $E(B-V)$), the authors rescaled one of the three structural components for any given map pixel to reproduce
the FIR flux.
The component, which is chosen for this rescaling, is typically that which needs the least fractional change in its column density to
account for the FIR residual, though the spiral arms are preferentially chosen near the Galactic midplane
and the disc component is chosen at high Galactic latitudes.

Finally, the authors constructed a set of 3D rectangular grids of $A_\mathrm{V}$. 
The large-scale grid with $200\times200\times20$~pc cells along the $X$, $Y$, and $Z$ axes cover the entire Galactic disc,
while the small-scale grid with smaller cells describes extinction near the Sun in greater detail.

\subsection{The Chen et al. (2014) map}

\citet[][hereafter CLY]{chen}
\footnote{\url{http://lamost973.pku.edu.cn/site/data}} used 2MASS and {\it WISE} photometry, as well as optical observations of the Xuyi 1.04/1.20~m 
Schmidt Telescope (XSTPS-GAC) in order to produce a 3D extinction map in the $r$ band.
The map covers an area of over 6000~deg$^2$ around the Galactic anticentre ($140^{\circ}<l<240^{\circ}$, $-60^{\circ}<b<40^{\circ}$)
with a spatial angular resolution (depending on latitude) between 3 and 9~arcmin.

Firstly, \citet{chen} defined a reference sample of zero-extinction stars by imposing the following criteria:
photometric errors in all the bands are smaller than $0.05$~mag
and the line-of-sight $A_\mathrm{r}$ extinction is less than $0.075$~mag
(i.e. $E(B-V)<0.028$~mag from the SFD map, taking into account $A_\mathrm{V}=1.172\,A_\mathrm{r}$, $E(B-V)=A_\mathrm{V}/3.1$, thus
\begin{equation}
\label{eqchen}
E(B-V)=0.378\,A_\mathrm{r}
\end{equation}
from the extinction law of \citet{yuan} used by the authors).
This selection produces a reference sample of 132,316 stars used to generate the standard SED library.
To create the map, a SED fitting was applied to the sample of about 13 million stars with the best photometry.

The authors used a sliding window of width 450~pc with a step of 150~pc to obtain a median value of $A_\mathrm{r}$
for each distance bin of 150~pc.
The CLY map spans a distance range from 0 to 4~kpc. However, at distances beyond 3~kpc it becomes less reliable
due to less numbers of stars with high-quality photometry available for estimating extinction.

To calculate $E(B-V)$ of a TGAS star from the CLY map, we select the line of sight nearest to that star, and interpolate the extinction
between the estimates for the two nearest distances along that line of sight.

\subsection{RAVE DR5 estimates}
\label{RAVE5}

\citet[][hereafter KKS]{kunder}\footnote{\url{https://www.rave-survey.org/project/}}
processed the data of the Radial Velocity Experiment (RAVE) survey \citep{binney} of spectra of
nearly 500,000 stars which have 2MASS photometry including 215,590 TGAS stars.
So far, the KKS is the only direct measurement of the extinction for a large number of the TGAS stars.
However, the majority of stars in the RAVE survey are middle-type main sequence stars at middle and high latitudes of the southern Galactic 
hemisphere ($b<-20^{\circ}$).
For O--F main sequence stars younger than 3 Gyr, which are considered in this study, the KKS sample is far from complete.

The authors made the first estimate of the distance to each star under the assumption $A_\mathrm{V}=0$.
Then they used this distance to put the star in the space where some physics would be applied
(luminosity, color, chemical composition, etc.), including evaluation of an $A_\mathrm{V}$ prior for this distance.
With this prior, the mutual processing of all available data continued for better parameters, including $A_\mathrm{V}$ for the stars.

The $A_\mathrm{V}$ (and $E(B-V)=A_\mathrm{V}/3.1$) prior was set using the reddening to infinity from the SFD map.
In addition, the authors estimated and took into account the SFD systematic errors at large reddening,
correcting the prior for this effect by the equation:
\begin{equation}
\label{eqbinney}
y=x\,(0.6+0.2\,(1-\tanh((E(B-V)-0.15)/0.3)))\,,
\end{equation}
where $x$ is original $E(B-V)$ from the SFD map and $y$ is corrected $E(B-V)$.
Finally, the authors concluded that the data sufficed to shift the recovered values away from the SFD-based prior. 
In particular, it follows from the fact that the final reddening for {\it Hipparcos} stars embedded in
the dust layer is, on average, lower than the SFD reddening to infinity along the same lines of sight.

\subsection{The Gontcharov (2010--2017) map}
\label{Gont_map}

\citet{map} analyzed the distribution of 70 million 2MASS stars with accurate photometry (better than $0.05$~mag) in the $(J-K_s)-K_s$ diagram.
One of the maxima of this distribution contains main sequence turn-off stars, i.e. F-type dwarfs, subdwarfs, and subgiants
with the mean absolute magnitude $\overline{M_\mathrm{K_s}}\approx3$~mag (other two maxima contain the clump giants and M dwarfs).
The shift of this maximum (i.e. $mode(J-K_s)$) toward larger $(J-K_s)$ with increasing $K_s$ reflects the reddening of these stars with the increasing 
distance $R$.
As the result, in every spatial cell the mean distance $\overline{R}$ and the mean magnitude $\overline{K_s}$ are proportional to each other:
\begin{equation}
\label{gon1}
\overline{R}=10^{(\overline{K_s}-\overline{M_\mathrm{K_s}}+5-\overline{A_\mathrm{K_s}})/5}.
\end{equation}
Hence, by dropping the turn-off stars into $K_s$, $l$, and $b$ cells one puts them into 3D spatial cells.
For every cell, the reddening is calculated with respect to a typical de-reddened color of the turn-off stars:
\begin{equation}
\label{ejk}
E(J-K_s)=mode(J-K_s)-(J-K_s)_0\,.
\end{equation}
However, $(J-K_s)_0$ may have natural spatial variations.
Therefore, in fact, two 3D maps were created: for $mode(J-K_s)$ and for $(J-K_s)_0$.
Their difference is the final 3D reddening map.

In the first version of the map, \citet{map} accepted $\overline{M_\mathrm{K_s}}=2.5$~mag and $A_\mathrm{K_s}=0$~mag.
The first version presented the reddening $E(J-K_s)$ with the precision of $0.03$~mag in the spatial cells of $100\times100\times100$~pc
along the rectangular Galactic coordinates $X$, $Y$, and $Z$ within 1600~pc from the Sun.

\citet{av} used the same 2MASS photometric data reduced by the same method as in the first version, but applied moving averaging to increase
the spatial resolution of the map.
The reddening $E(J-K_s)$ and $E(B-V)$ were calculated by assuming the relation~(\ref{arebv}) and the extinction law of \citet{rl}:
\begin{equation}
\label{rllaw}
E(B-V)=1.9\,E(J-K_s)
\end{equation}
for the same spatial cells of $100\times100\times100$~pc as those in the first version, but each cell was moved for averaging by 50~pc
(instead of 100~pc) in each of the coordinates $X$, $Y$, or $Z$.
Thus, the moving averaging yields the second version of the map with the spatial resolution of 50~pc.

The product of this second version of the 3D reddening map and the 3D map of spatial variations of $R_\mathrm{V}$ constructed by
\citet{rv} allowed \citet{av} to create a 3D map of $A_\mathrm{V}=E(B-V)\,R_\mathrm{V}$.

Recently, \citet[][hereafter G17]{g17} presented the third, improved version of the map based on the same data which were reprocessed by a similar but 
a more sophisticated method.
Here we present a brief description of this third version.

Following the extinction law of \citet{cardelli} , the relations~(\ref{arebv}) and (\ref{rllaw}) are replaced by the 
relations~(\ref{arebv}), (\ref{bvjk}) and (\ref{akav}) in the new version of the map.

An analysis of the distribution of the TGAS stars with the most precise parallaxes in the HR diagram gives 
$0.2<M_\mathrm{K_s}+A_\mathrm{K_s}<3.6$~mag for turn-off stars.
Together with the interval $5<K_s<14$~mag, where the 2MASS photometry is precise enough (better than 0.05~mag, the median error is 0.02~mag), it provides
the complete sample of the turn-off stars with the precise photometry within 1200~pc from the Sun.
Due to the decrease of the stellar mass density far away from the Galactic midplane, an additional limit $|Z|<600$~pc is applied.
Thus, the space where the method would give a reliable map is limited by $R<1200$ and $|Z|<600$~pc.
This space is divided into 630,109 cubic cells of $20\times20\times20$~pc in order to derive the reddening $E(J-K_s)$ in every cell.

Moreover, the reddening is calculated by the same method and for the same space, but depending on the spherical coordinates $R$, $l$, and $b$
for the cells of 20~pc $\times 1^{\circ} \times 1^{\circ}$. Such an approach allows G17 to analyze variations of the reddening and de-reddened
color along the lines of sight. Also, it allows G17 to construct a map of differential reddening inside every cell
in addition to the map of the cumulative reddening from the observer to the cell.

In some cases the reddening along a line of sight decreases with the distance at some short interval of distance.
Such local depressions reach $\Delta E(B-V)=-0.04$~mag.
\citet{map} concluded that this effect is related to the limited and moderate angular resolution of the map.
For a spatial cell slightly larger than a dust cloud within it, stars located beyond the cloud but outside its projection on to the celestial
sphere are observed in the part of the cell unobscured by the cloud. Being outside the cloud, usually they have a lower reddening than
the stars in and behind the cloud.
Therefore, the $R - E(J-K_s)$ relation for such a line of sight shows a maximum reddening in the cell with the cloud, namely, 
at the distance of the far side of the cloud.
Then, in the next cell at larger $R$ the reddening decreases due to the contribution of the stars located farther than the cloud but outside
its projection on to the celestial sphere.
This decrease by no means implies that the reddening behind the cloud is lower than in it.
G17 provided a detailed inspection of the cases where the reddening decreases with increasing distance and showed that, indeed,
as was assumed earlier, all of them appear to be behind some dense dust clouds.
In addition, two other possible reasons of such decrease of reddening/extinction with the distance are 
the result of errors of the stellar distances and variations of the stellar content of the sample.
Indeed, the former reason may cause an effect when some stars with higher reddening/extinction look closer, whereas some stars with lower 
reddening/extinction seem to be farther away.
The latter reason can add more intrinsically blue or red stars to the spatial cell.

\citet{sale} and \citet{green} have developed an approach to take into account small-scale fluctuations of the dust medium in order to ensure 
that the reddening does not decrease with the distance in reddening maps.
However, their solution is not conclusive as the stellar distances are quite uncertain and the stellar content in a cell may vary significantly (especially, in 
case of G17 with only few tens of stars in the cell on average).
Therefore, G17 has not treated the cases of the reddening depression because 
(a) the influence of uncertain distances, content variations, and medium fluctuations cannot be separated in G17, 
(b) cases when the reddening decreases are quite rare, and 
(c) they contribute little to the total errors of the G17 map.

Unlike the previous versions of the map with the fixed $\overline{M_\mathrm{K_s}}$ and the neglected $A_\mathrm{K_s}$,
in the new version some parameters of every star with $0.2<(J-K_s)<0.8$~mag are mutually refined iteratively solving the set of the equations:
\begin{equation}
\label{sist}
\left\{
\begin{aligned}
A_\mathrm{V}=f_1(R, l, b)\\
A_\mathrm{K_s}=0.117\,A_\mathrm{V}\\
R=10^{(K_s-A_\mathrm{K_s}-M_\mathrm{K_s}+5)/5}\\
(J-K_s)_0=f_2(R, b)\\
M_\mathrm{K_s}=f_3\,(J-K_s)_0\,,
\end{aligned}
\right.
\end{equation}
where $f_1$, $f_2$, and $f_3$ are some functions described below.

This approach is similar to the one applied by \citet{ob} for the study of the complete sample of 20,514 OB stars from the {\it Tycho-2}.
That sample largely overlaps with the sample of the TGAS O--F main sequence stars used in the current study.
Therefore, the consistent distance, reddening, extinction, absolute magnitude, de-reddened color, age, and spatial distribution,
obtained by \citet{ob} from a system of equations similar to~(\ref{sist}) for those OB stars, attest to the reliability of the current approach.

The interval $0.2<(J-K_s)<0.8$~mag always contains the required maximum of the distribution of the turn-off stars in the $(J-K_s)-K_s$ diagram,
together with the wings of this distribution which are necessary  to define the maximum.
An advantage of this method is that there is no selection of stars and no corresponding mistakes and errors.

In order to suppress the influence of the distance uncertainty, short-term spatial variations of the reddening and the
cases of its decreasing with distance, every star is taken into account in 19 spatial cells:
one with its $X$, $Y$ and $Z$ coordinates and the remaining 18 cells that share a face or edge with the cell under consideration.
Thus, in fact, G17 applies a spherical window with the smoothing radius of 28.3~pc.
This does not contradict the determination of the unique $R$ for every star because its accuracy is worse than 28.3~pc.

A similar approach is applied for the spatial cells of $R$, $l$, and $b$. But the linear size of such a spatial cell varies due to its fixed angular size.
Therefore, G17 varies the number of the averaged spatial cells in order to obtain not less than 5 stars in every cell in the 4D space $R-l-b-mode(J-K_s)$.

As we solve the system~(\ref{sist}), during every iteration we re-sort the stars into spatial cells and also re-sort them within every spatial cell
into 60 $J-K_s$ cells of 0.01~mag in the range $0.2<(J-K_s)<0.8$~mag.
Thus, in every spatial cell we find the mode of the stellar distribution on $J-K_s$ with the precision of 0.01~mag.
Consequently, during every iteration the mode of the distribution of the turn-off stars may shift both in space as along $J-K_s$ due to
changing $R$, $X$, $Y$, $Z$, $(J-K_s)_0$, and $J-K_s$.
For all spatial cells under consideration the iterations converge to a solution after less than 14 iterations.
As for the previous versions of the map, the reddening to the spatial cell is calculated following the equation~(\ref{ejk}) for
the $J-K_s$ cell which corresponds to the mode of the turn-off stars distribution along $J-K_s$.

The 3D model of the spatial variations of the extinction $A_\mathrm{V}$, as a function of $R$, $l$, and $b$ \citep{gould, av}
(see Sect.~\ref{Gont_model}), was used for the first approximation of the function $f_1$ in the system~(\ref{sist}) and also as a constraint on $f_1$ 
in the following approximations. This constraint is:
\begin{equation}
\label{constr1}
E(J-K_s)<\min(0.8;(1.08\,A_\mathrm{V}+0.89)/3.1/1.655)
\end{equation}
for the hemisphere of the Galactic anticentre ($X<0$) and
\begin{equation}
\label{constr2}
E(J-K_s)<\min(0.8;(0.71\,A_\mathrm{V}+0.79)/3.1/1.655)
\end{equation}
for the hemisphere of the Galactic centre ($X>0$), where $A_\mathrm{V}$ is the extinction from the 3D model.
This is a rather weak constraint which allows G17 to recognize a descending wing after the maximum of the distribution of the stars along $(J-K_s)$ in every 
spatial cell.
As a result of the solution of the system~(\ref{sist}), the estimates given by the function $f_1$ significantly deviate from the ones given by the 3D model.
Taking into account the relations~(\ref{arebv}) and (\ref{bvjk}), the final function $f_1$, being tabulated as a function of $R$, $l$, $b$ or $X$, $Y$, $Z$,
is the final reddening map.

\begin{figure}
\includegraphics{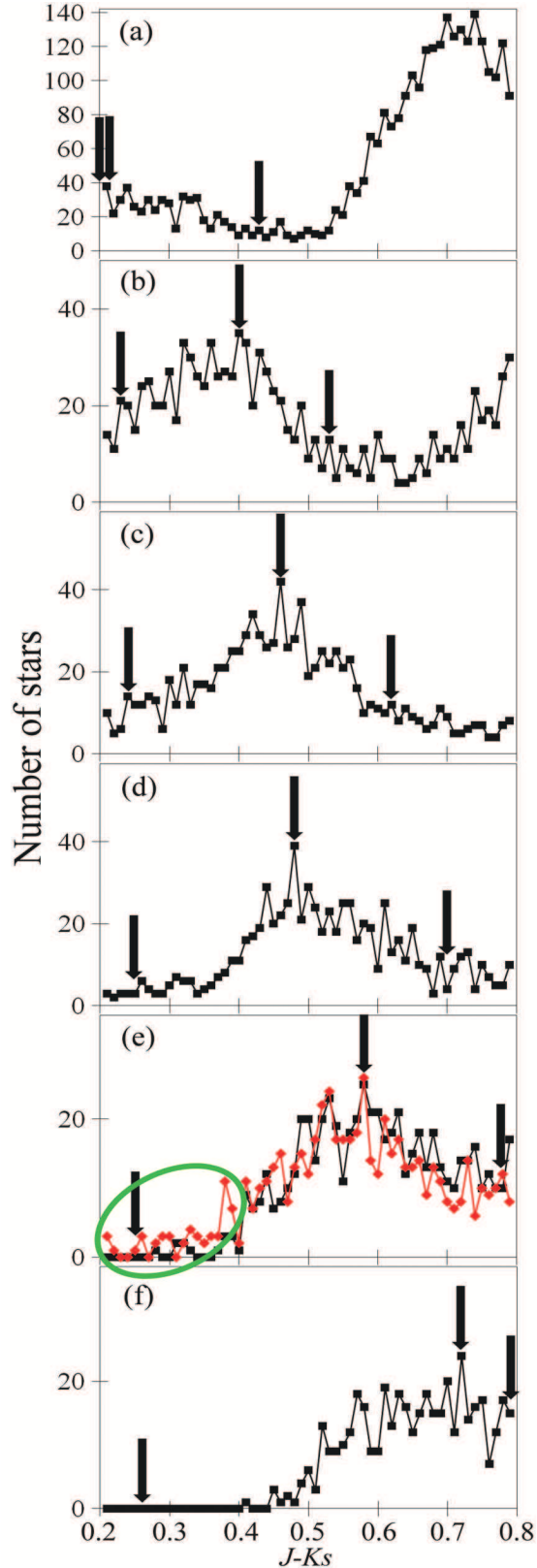}
\caption{The distribution of the 2MASS stars in the $J-K_s$ cells of the spatial cells in the direction
$l=143.13^{\circ}$, $b=0^{\circ}$ at (a) $R=100$, (b) 300, (c) 500, (d) 700, (e) 900 and (f) 1100 pc -- black squares.
The same for $l=143.88^{\circ}$, $b=+1.25^{\circ}$, $R=916$ pc -- red (grey) diamonds.
Some reddened stars of the OB association Cam OB1 -- green ellipse.
The left, central and right arrows show, respectively, $(J-K_s)_0$, $mode(J-K_s)$, and a limiting $(J-K_s)$ to which $mode(J-K_s)$ was searched.
}
\label{jkncamob}
\end{figure}

This approach is illustrated in Fig.~\ref{jkncamob} with an example of the distribution of 2MASS stars in $J-K_s$ cells of 0.01~mag width
for spatial cells of $20\times20\times20$~pc in the direction
$l=143.13^{\circ}$, $b=0^{\circ}$ at the distances: (a) $R=100$, (b) 300, (c) 500, (d) 700, (e) 900 and (f) 1100 pc shown by the black squares.
The left, central and right arrows in every plot show, respectively, $(J-K_s)_0$, $mode(J-K_s)$, and the limiting $(J-K_s)$ to which $mode(J-K_s)$ was searched.
A shift of the arrows with the distance is evident. It follows the increasing reddening $E(J-K_s)= 0.01$, 0.17, 0.22, 0.23, 0.33, 0.46~mag 
(distance between the left and central arrows in the plots)
for $R=100$, 300, 500, 700, 900 and 1100 pc, respectively. 
Also, Fig.~\ref{jkncamob} shows that the determination of the $mode(J-K_s)$ is quite robust, with the accuracy of 0.01~mag.
The second maximum of the distribution of the stars, which is clear for $R=100$ and 200 pc at $J-K_s>0.6$~mag, is due to the presence of the clump giants.
To remove them, the constraint~(\ref{constr1}) or (\ref{constr2}) was adjusted.

It is seen in Fig.~\ref{jkncamob} that for large distances the reddening is so high that no stars bluer than $J-K_s\approx0.4$~mag are found (except for some 
cases when a large well-reddened OB association or open cluster is encountered).
As an example, the distribution of stars in the cell $l=143.88^{\circ}$, $b=+1.25^{\circ}$, $R=916$ pc is shown by the red (grey) diamonds with
the reddened stars of the OB association Cam~OB1 circled by the green ellipse.
Being close to each other, the cells $l=143.13^{\circ}$, $b=0^{\circ}$, $R=900$ pc and $l=143.88^{\circ}$, $b=+1.25^{\circ}$, $R=916$ pc,
as is evident from the plot (e), show nearly the same distribution of the stars for $J-K_s>0.4$~mag, but the latter contains some 35 extra stars of the
Cam~OB1 association at $J-K_s<0.4$~mag. 
Indeed, given $E(J-K_s)=0.33$~mag and $0.2<J-K_s<0.4$~mag, their de-reddened color is $-0.13<(J-K_s)_0<0.07$~mag, in agreement with the color of B and A stars.
G17 found such extra stars for the Cam~OB1 area of the sky only for the distance range of $900<R<1000$ pc which can be assigned for this association.
This distance estimate agrees with the one from \citet{dezeeuw}. 
Thus, with this approach we are able to estimate the distance, reddening, and extinction to some large well-reddened ($E(J-K_s)>0.2$~mag) 
OB associations and open clusters as a by-product.

The function $f_3$ in the system~(\ref{sist}) describes the main sequence in the HR diagram, except for K and M dwarfs.
By use of the best data for the {\it Gaia} DR1 TGAS stars in the range $0.1<(J-K_s)<0.45$~mag, G17 calculated it by the method of least squares:
\begin{equation}
\label{jk0mks}
M_\mathrm{K_s}=0.22+9.43\,(J-K_s)_0\,-5.57\,(J-K_s)_0^2\,.
\end{equation}
The coefficients of this relation were calculated for various samples of the TGAS stars: with and without the giants, and for various limitations
of color. Also, G17 tried to replace the equation~(\ref{jk0mks}) by some polynomials of degree up to 4.
However, these refinements of the approximation change the final values of $E(J-K_s)$ by less than 0.01~mag in most spatial cells.
Furthermore, the coefficients of the relation~(\ref{jk0mks}) were calculated using the Monte Carlo simulation discussed below, and it appeared that they differ insignificantly.
Moreover, the equation~(\ref{jk0mks}) was fitted to the {\it Hipparcos} stars with the most accurate $\varpi$.
It is shifted with respect to the TGAS's equation by $\Delta M_\mathrm{K_s}=0.12$~mag in the sense `TGAS minus Hipparcos'.
This means that the TGAS main sequence is less luminous due to the selection effect in {\it Hipparcos} in favour of brighter stars.

The function $f_2$ in the system~(\ref{sist}) presents some spatial variations of the de-reddened color $(J-K_s)_0$ of the turn-off
stars. In fact, this function is the 3D analytical map of these variations.

The basic assumption of this method is that the mean de-reddened color $\overline{(J-K_s)_0}$ and the
related mean absolute magnitude $\overline{M_\mathrm{K_s}}$ of the stars constituting the maximum of the distribution
in the $(J-K_s)-K_s$ diagram are constant or predictable in some space.
It is equivalent to the assumptions that (a) the stellar population changes slightly and smoothly within this space
and
(b) the 2MASS catalogue is quite complete in this space to the limit of $K_s<14$~mag.
The latter is important because the mean de-reddened color $\overline{(J-K_s)_0}$ may be biased in dependence on $R$ due to selection effects.
These assumptions constrain the space where this method can be applied, as was previously discussed.

\citet{map} tested these assumptions for the {\it Hipparcos} stars as well as in the Monte Carlo simulation of the 2MASS catalogue content based on the BMG \citep{bmg1}
and PARSEC theoretical isochrones \citep{girardi}, as applied to the space within 1.6~kpc from the Sun and with the limitation $K_s<14$~mag.
This simulation is repeated for the third version of the map, taking into account the new version of the BMG \citep{bmg2} and new
PARSEC isochrones \citep{bressan}. The simulation justifies using the relation~(\ref{jk0mks}) and provides a slightly different function $f_2$
than in the previous versions of the map. 

For the previous versions, the simulation and fitting to the real data constitute:
\begin{equation}
\label{gon3}
(J-K_s)_0=0.09\,\sin|b|+0.00004\,R\,,
\end{equation}
where $R$ is expressed in pc.
The coefficients of this equation were fitted by minimizing the reddening for all lines of sight assuming that the reddening $E(J-K_s)$ satisfies the following conditions: 
\begin{itemize}
\item
smoothly tends to 0 when $R$ tends to 0,
\item
is negative in a statistically insignificant number of cells taking into account the mean precision of the photometry used,
\item
decreases only locally, i.e. for every line of sight it can be approximated by a smooth nondecreasing function.
\end{itemize}

Taking into account the high precision of the photometry used (for all the space we consider in this study, the median precision in every band is 0.02~mag)
and quite large number of stars in a typical cell (on average, 21 stars are in the cell of the 4D $R-l-b-mode(J-K_s)$ space),
the second condition means that the negative reddening is 
allowed only for distances $R<40$ pc, where the 2MASS photometry is saturated.

The simulation made for the new version of the map allows G17 to conclude that $(J-K_s)_0$ depends on $\log(R)$ rather than on $R$ in the equation~(\ref{gon3}).
Consequently, the smooth non-decreasing function mentioned earlier in the third condition is a logarithmic function.
Therefore, instead of the equation~(\ref{gon3}) the function $f_2$ accepted for the new version of the map becomes:
\begin{equation}
\label{gon4}
(J-K_s)_0=0.10+0.09\,\sin|b|+0.0518\,\log(R)\,,
\end{equation}
where $R$ is in pc.

\begin{figure}
\includegraphics{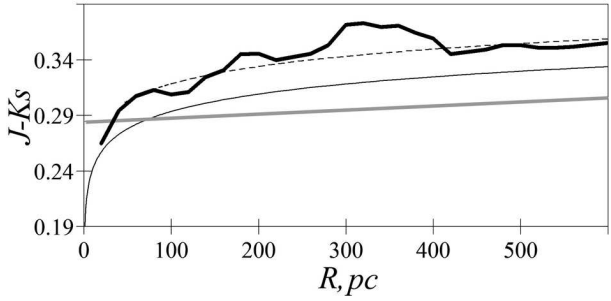}
\caption{$(J-K_s)$ in dependence on $R$ in the direction of SGP -- black thick curve;
an approximation of this dependence by a logarithm -- black dotted line;
the relation~(\ref{gon4}) -- black thin curve;
the relation~(\ref{gon3}) for the previous version of this map with the law~(\ref{rllaw}) replaced by the law~(\ref{bvjk}) -- grey thick line. 
}
\label{sgp}
\end{figure}

As for the previous versions of the map, the coefficients of the equation~(\ref{gon4}) are fitted by minimizing the reddening for all lines of sight under
the same conditions as earlier.
Following them, the strongest constraints are applied, in fact, for the lines of sight with minimal reddening, i.e. in the celestial areas 
listed by SFD in their table~5, including the areas around the Galactic poles.
The minimal coefficient at $\log(R)$ in the equation~(\ref{gon4}) is found for the area around the southern Galactic pole.
Fig.~\ref{sgp} shows the estimates in this direction.
The observed variations of $(J-K_s)$ depending on distance are shown by the black thick curve,
an approximation to this dependence by the equation $J-K_s=0.215+0.0518\,\log(R)$ ($R$ in pc) -- by the black dotted line,
the relation~(\ref{gon4}) -- by the black thin curve, and
the relation~(\ref{gon3}) for the previous version of this map with the law~(\ref{rllaw}) changed by the law~(\ref{bvjk}) -- by the grey thick line.
One can see that for $R>200$ pc the grey line and black thin curve are approximately parallel to each other. Their difference is mainly due to the change
of the extinction law, whereas the replacement of $R$ in the equation~(\ref{gon3}) by $\log(R)$ in the equation~(\ref{gon4}) is important only for $R<200$ pc.
The change of the extinction law means that for $R>75$ pc the new version of the map gives higher $(J-K_s)_0$ and, consequently, lower $E(J-K_s)$ than
the previous versions.
Fig.~\ref{sgp} shows unambiguously the determination of the coefficients $0.1$ and $0.518$ in the equation~(\ref{gon4}) as, respectively, the vertical shift
and tilt of the black thin curve.
A lower tilt would give a surplus systematic increase of $E(J-K_s)$ (the difference between the solid black lines) at large distances,
a higher tilt would give a meaningless decrease of $E(J-K_s)$ with the distance.
An upper shift would give negative $E(J-K_s)$ for small distances,
whereas a downward shift would give $E(J-K_s)$ far from zero for small distances.

After analyzing the areas with minimal reddening, the zero-point 0.1 and the coefficient 0.0518 at $\log(R)$ were fixed for further fitting of the 
coefficient at $\sin|b|$ to the data at all $b$. The precision of the coefficients in the equation~(\ref{gon4}) is the unit of the last digit.

The precision of the map can be estimated from the scatter of its values in large regions of space far from the Galactic midplane.
Such scatter includes natural variations of the dust medium together with some variations due to random errors of the map.
G17 found this scatter to be $\sigma(E(J-K_s))<0.012$~mag. It should be considered as the upper limit of the precision of the map.
The precision should be the same for the whole space under consideration because the precision of the 2MASS photometry is constant here (the median is 0.02~mag).
Also, it is important to stress that the value of $J-K_s$ in every spatial cell, which is the key one for the equation~(\ref{ejk}),
is determined not by some averaging of the values for individual stars, but as the mode of the distribution of the stars in the cell.
The precision of the mode is exactly the partitioning step, i.e. 0.01~mag, as mentioned before.

The systematic errors of the map can be larger. They include the uncertainty of the coefficients in the equations~(\ref{jk0mks}) and (\ref{gon4}).
This translates into $\sigma(E(J-K_s))<0.02$~mag based on the precision of the initial data and the results of the simulation.
Also the systematic uncertainty includes that of the coefficients~(\ref{arebv}), (\ref{bvjk}), and (\ref{akav}) of the accepted extinction law.
Due to poorly known systematic spatial variations of the law, their contribution to the uncertainty cannot be estimated.
Finally, with some caution, the systematic accuracy of the map can be evaluated as $\sigma(E(J-K_s))=0.025$~mag, or $\sigma(E(B-V))=0.04$~mag.

We use trilinear interpolation of the G17 map to obtain the reddening for the selected TGAS stars.

\subsection{The Arenou, Grenon \& Gomez (1992) model}
\label{arenou_map}

Since the middle of the 20$^{th}$ century there have been attempts to present the Galaxy reddening or extinction as an analytical function of
the Galactic coordinates $l$, $b$, and $R$.
They are referred to as analytical 3D models. Based on them, the reddening can be estimated for any star.
 
All available precise spectral and photometric data for more than 42,000 stars were processed by 
\citet[][hereafter AGG]{arenou} and presented as an analytical 3D model of the interstellar
extinction $A_\mathrm{V}$ within the distance of about 1~kpc from the Sun.
The sky was divided into 199 cells by Galactic coordinates.
In each cell, the following quadratic relation was adopted up to a distance limit of the dust layer:
$A_\mathrm{V}=k_1\,R+k_2\,R^2$\,,
where $k_1$ and $k_2$ are some coefficients found empirically for every cell.
Out of the layer, $A_\mathrm{V}$ remains constant and identical to $A_\mathrm{V}$ at the edge of the layer
(at higher latitudes) or slowly growing following a linear regression relation (at lower latitudes).
Some constraints on $A_\mathrm{V}$ were taken into account:
$A_\mathrm{V}<0.1$~mag for $|b|\ge60^{\circ}$, $A_\mathrm{V}<1.2$~mag for $45^{\circ}\le|b|<60^{\circ}$, and
$A_\mathrm{V}<3$~mag for $|b|<45^{\circ}$.
The average relative error of $A_\mathrm{V}$ in the whole sky is about 35~per cent. However it significantly varies from cell to cell.

This model has some features:
\begin{enumerate}
\item
They did not present any physical explanation for the observed systematic spatial variations of $A_\mathrm{V}$.
\item
They used the single functional dependence of $A_\mathrm{V}$ on $R$ inside every area in the sky,
i.e. these area dependencies are inconsistent with each other. This means that $A_\mathrm{V}$ may significantly differ
even for close lines of sight.  
\item
As any 3D extinction model, it provides a general trend of the extinction and cannot take
into account local irregularities of the dust medium.
\item
The majority of the analyzed stars are O--F main sequence stars within 600~pc from the Sun and near the Galactic midplane
resulting in lower accuracy of $A_\mathrm{V}$ at large distances and high latitudes.
\end{enumerate}

\subsection{The Gontcharov (2012) model}
\label{Gont_model}

\citet{gould} proposed a 3D analytical model of the spatial variations of the extinction as an
alternative to the previously discussed model of \citet{arenou}.
This model describes the extinction $A_\mathrm{V}$ at every point within at least
the first kiloparsec by the sum of the extinctions in two dust layers:
along the Galactic equatorial midplane and along the midplane of the Gould Belt.

The SFD map and other sources show that the dust is distributed in the solar neighbourhood not only along the Galactic midplane
but also along the Gould Belt, whose general description was given in e.g. \citet{perr} (pp. 324--328) and
\citet{bobgould} (also, see references therein).
The Gould Belt contains dust clouds, starforming regions, young stars, their associations, and young clusters. 
\citet{taylor} were among the first who mentioned the Gould Belt as a container of dust entering the Solar system.
\citet{vergely} mentioned the importance of the interstellar extinction in the Gould Belt.

In the model, the two dust layers intersect at an angle $\gamma$.
The axis of intersection between the layers is turned relative to the $Y$ axis through an angle $\lambda_0$. 
The Gould Belt layer has a finite radius $R_\mathrm{limit}$ and centered relative to the Sun.
This radius can be calculated as a free parameter of the model or taken fixed (its estimates $R_{limit}<600$~pc look reasonable).
The calculation of extinction in the Belt should be restricted to this radius.

Within every layer the extinction follows the barometric law and depends on the cylindrical coordinates:
\begin{equation}
\label{aeq2}
A(R, l, Z)=(A_{0}+A_{1}\sin(l+A_{2}))\,R\,(1-\mathrm{e}^{-|Z-Z_{0}|/Z_{A}})\,Z_\mathrm{A}/|Z-Z_{0}|\,.
\end{equation}
This is the equatorial layer extinction, and
\begin{equation}
\label{ago2}
A(R,\lambda,\zeta)=(\Lambda_{0}+\Lambda_{1}\sin(2\lambda+\Lambda_{2}))\min(R,R_\mathrm{limit})(1-\mathrm{e}^{-|\zeta-\zeta_{0}|/\zeta_\mathrm{A}})\,\zeta_\mathrm{A}/|\zeta-\zeta_{0}|
\end{equation}
is the Gould Belt extinction.
Here $\zeta=R\,\sin(\beta)$ is the shift of a point in the coordinate system of the Gould Belt
perpendicular to the midplane of the Belt, whereas
$\lambda$ and $\beta$ are the longitude and the latitude of the point in the coordinate system of the
Gould Belt calculated as
\begin{equation}
\label{equ1}
\sin(\beta)=\cos(\gamma)\sin(b)-\sin(\gamma)\cos(b)\cos(l)
\end{equation}
\begin{equation}
\label{equ2}
\tan(\lambda-\lambda_{0})=\cos(b)\sin(l)/(\sin(\gamma)\sin(b)+\cos(\gamma)\cos(b)\cos(l)).
\end{equation}
$A_0$, $A_1$, and $A_2$ are the free term, the extinction amplitude,
and the phase in the sinusoidal dependence on $l$; whereas
$\Lambda_0$, $\Lambda_1$, and $\Lambda_2$ are the same parameters in the dependence on $2\lambda$.
The assumption that the extinction in the Gould Belt has two maxima in the dependence on $\lambda$ was confirmed by the following data fitting.

In fitting the model, \citet{gould} used individual extinctions for tens of thousands of stars from three catalogs, whereas \citet{av} used
the 3D map of $A_\mathrm{V}$ constructed in that paper and based on the second version of Gontcharov's map discussed earlier in our paper
(see Sect.~\ref{Gont_map}).
Consequently, the model is slightly related to the third version of the G17 map.
The best up-to-date fitting to the model (hereafter the G12 model) is the extinctions $A_\mathrm{V}$ as the sum of the extinctions in the equatorial
and Gould Belt layers, respectively:
\begin{equation}
\label{model1}
(1.2+0.3\,\sin(l+55^{\circ}))R(1-\mathrm{e}^{-|Z+0.01|/0.07})0.07/|Z+0.01|\,,
\end{equation}
\begin{equation}
\label{model2}
(1.2+1.1\,\sin(2\lambda+130^{\circ}))\min(0.6,R)(1-\mathrm{e}^{-|\zeta|/0.05})0.05/|\zeta|\,,
\end{equation}
where the angular values are expressed in degrees and the distances are given in kpc.

\subsection{Inappropriate estimates}
\label{wrong}

Every data source of reddening/extinction estimates is reliable for some defined volume of space. Such a volume is
usually close to the volume declared by the authors of this source. Since in this study we consider stellar distances $R<280$~pc, we do not include some 
sources which cannot provide us with reliable estimates of the reddening/extinction.
For example, we do not consider:
\begin{itemize}
\item The 3D map from \citet{marshall} which covers only a small part of the sky ($|b|<10^{\circ}$, $|l|<100^{\circ}$) and
loses most information within the first kiloparsec, as noted by its authors.
\item Extinction estimates from \citet{berry}\footnote{\url{http://www.astro.washington.edu/users/ivezic/sdss/catalogs/tomoIV/}}
which cover a part of the sky and give unreliable results within the nearby space under consideration due to saturated photometry, 
as noted by their authors.
\item The 3D map from \citet{sale}\footnote{\url{http://www.iphas.org/extinction/}} which covers only a small part of the sky
($|b|<5^{\circ}$, $30^{\circ}<l<215^{\circ}$) and has the largest 
uncertainties within the first kiloparsec, where the number of stars is limited by the IPHAS bright magnitude limit of $r\approx13$~mag, as stated by 
the authors.
\item The 3D map from \citet{green}\footnote{\url{http://argonaut.skymaps.info/}} which covers only three-quarters of the sky  ($\delta>-30^{\circ}$), uses 
Pan-STARRS and 2MASS photometry for rather faint stars and, consequently, has no reliable results for a few hundred parsecs near the Sun 
(here it follows the estimates of the DCL's smooth disc component), as stated by the authors.
\end{itemize}
It is worth noting, that these data sources might provide reliable estimates of reddening/extinction far from the Sun.

Despite the warning by their authors, these data sources have been quite popular for reddening and extinction estimates within the local space.
For example, such inappropriate estimates of the reddening/extinction within the local space may affect the results of
\citet{macdonald}, \citet{obermeier}, \citet{kadofong}, \citet{lund}, \citet{vaneck}, \citet{davies}, \citet{munn}, \citet{grobela},
\citet{pety}, and \citet{narayan}.
Moreover, in some of these studies their authors admit that the reddening/extinction estimates may be inappropriate. 
But, unfortunately, they pay no attention to alternative estimates.
One of the tasks of our study is to recommend such alternative estimates.

\section{Results}
\label{results}

To estimate the reddening and extinction outside the dust layer, the SFD and PLA maps are sufficient.
Yet, an issue arises: are the TGAS stars located inside or outside the dust layer?
Or, in other words, for proper estimates, should one reduce the reddening/extinction from infinity to the distance?
This leads to a practical definition of the dust layer edge at the current level of our sensitivity as a point where the extinction from the Sun to it is indistinguishable (at the given precision level) from the extinction to infinity.
Obviously, the higher the accuracy of data used, the higher our sensitivity to extinction, and, hence, the larger the distance from the Galactic midplane to the layer edge
defined in this way.
This does not mean, however, that there is no dust outside this edge 
(moreover, \citet{g2013, g2016} have shown that there is interstellar dust and related extinction law spatial variations up to 3 and even 25~kpc from 
the Galactic midplane).
This only means that at the current level of accuracy the reddening and extinction of any source outside this edge can be taken from a 2D map, 
without its laborious conversion to a 3D one.
Thus, from a practical point of view, we should count  the TGAS stars inside the layer as the ones with a detectable reddening difference
between the SFD or PLA estimates outside the layer and the estimates inside the one by PLA$_R$, SFD$_R$ and the remaining data sources.
We accept the value of the detectable reddening difference as $\Delta E(B-V)>0.03$~mag (or $\Delta A_\mathrm{V}>0.1$~mag).
This is an estimate of the accuracy of the best data sources we consider here.

It appears that the percentage of all the TGAS stars inside the layer does not depend on the relative error of the parallax
$\sigma(\varpi)/\varpi$ and has little variations from one data source to another.
For example, for $\sigma(\varpi)/\varpi<0.1$ 
67~per cent of all the TGAS stars have $\Delta E(B-V)>0.03$~mag for PLA minus PLA$_R$, 
72~per cent for PLA--DCL, 
59~per cent for PLA--G17, 
64~per cent for PLA--AGG;
for $\sigma(\varpi)/\varpi<0.2$ 61~per cent for PLA--PLA$_R$, 
66~per cent for PLA--DCL, 
65~per cent for PLA--G17, 
64~per cent for PLA--AGG.
Similar percentages are found for the differences with the SFD map.
Thus, we may conclude that the majority of the TGAS stars are located inside the dust layer.

For $\sigma(\varpi)/\varpi<0.5$ 
(it is just a selection of more distant stars, the accuracy of $\varpi$ is not important in this test)
the percentage of the TGAS stars with $\Delta E(B-V)>0.03$~mag is still high:
54~per cent for PLA--PLA$_R$, 
61~per cent for PLA--DCL, 
77~per cent for PLA--G17, 
68~per cent for PLA--AGG.
Therefore, a significant proportion of all {\it Gaia} stars might be embedded in the dust layer.
Any 2D map does not seem to be enough to precisely estimate their reddening and extinction.

We can estimate the $|Z|$ distance of the dust layer edge from the Galactic midplane (roughly half-thickness of the layer at the Sun)
and compare it with the layer scale height.
We find the difference $\Delta E(B-V)>0.03$~mag between the PLA and PLA$_R$ estimates up to about $|Z|=290$~pc with the layer scale height
$Z_\mathrm{A}=100$~pc, and up to about $|Z|=350$~pc with $Z_\mathrm{A}=140$~pc accepted in the BMG.
Thus, with precise data we can detect the dust at a $|Z|$ distance several times the dust layer scale height.
Taking into account the above-mentioned cut-off $R<280$~pc, we conclude that the selected stars are inside the dust layer.

The TGAS is quite complete only for the main sequence stars of spectral classes O--G and red giants.
The giants are few within $R<280$~pc and they are so luminous that many of them are saturated in IR surveys.
Thus, we have to use the main sequence stars of spectral classes O--F, far from the giant branch to avoid mixing.
It approximately corresponds to age of less than 3~Gyr.
To test the reddening/extinction data sources described in Sect.~\ref{maps}, we compare the location of this sample of the TGAS stars in
the HR diagram with the PARSEC and MIST \citep{mist}\footnote{\url{http://waps.cfa.harvard.edu/MIST/}} theoretical isochrones.
In this domain of the HR diagram the isochrones are well-defined and distinct in dependence on age and metallicity.
At the main sequence the difference between PARSEC and MIST isochrones of similar age and metallicity is at the level of the precision of 
photometry used: for example, $\Delta(G-H)_0<0.03$, $\Delta(G-W2)_0<0.03$, $\Delta(V_T-K_s)_0<0.03$~mag.
This means that the MIST isochrones almost coincide with the PARSEC ones in the scale of the subsequent figures with the HR diagram, and are not shown 
for clarity.
Yet, both isochrones are considered in the subsequent calculations.

\begin{figure}
\includegraphics{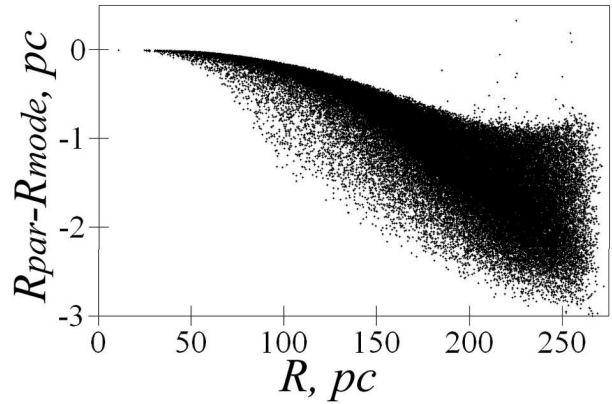}
\caption{Difference between the distances $R_{par}=1/\varpi$ and $R_{mode}$ calculated by \citet{bailer3} (parsecs) in dependence on $R_{par}$ (parsecs).
}
\label{rdiff}
\end{figure}

\begin{figure}
\includegraphics{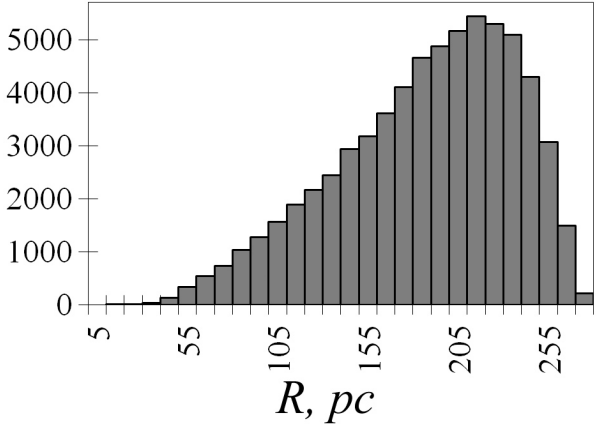}
\caption{Histogram of the distances for the stars under consideration.
}
\label{histr}
\end{figure}

Fig.~\ref{rdiff} shows the difference between the distances $R_{par}=1/\varpi$ and $R_{mode}$ calculated by \citet{bailer3} (in parsecs) in dependence 
on $R_{par}$ (in parsecs). This difference is mainly less than 1~per cent of the distance. It supposes that any bias introduced by the usage of one or another kind of 
distance is lower than 1~per cent and, consequently, is negligible with respect to the uncertainty of parallax $\sigma(\varpi)/\varpi<0.1$.
Therefore, $R_{par}$ and $R_{mode}$ lead to the same results. 
Fig.~\ref{histr} shows the histogram of the distances for about 60,000 stars under consideration.

The limitation $R<280$~pc for the O--F main sequence stars means that the photometry precision is better than $0.05$~mag only in the {\it Gaia} $G$ band,
$J$, $H$, $K_s$ bands from 2MASS and $W2$ band from WISE.
Also, the $V_\mathrm{T}$ band from {\it Tycho-2} can be used after the limitation $V_\mathrm{T}<10.5$~mag for an acceptable precision of photometry.
This limitation reduces the sample by one-third.
$B_\mathrm{T}$ photometry from {\it Tycho-2} and photometry in WISE $W3$ and $W4$ bands are precise only for bright stars, WISE $W1$ and all bands 
from APASS are saturated for many selected stars,
whereas SDSS photometry is mainly related to high latitude stars outside the dust layer.
Also, the extinction in the $W3$ band is rather uncertain due to the uncertainty of the extinction law:
for example, $A_\mathrm{W3}=0.002\,A_\mathrm{V}$ vs. $0.089\,A_\mathrm{V}$ following the extinction law of \citet{cardelli} and that of
\citet{wd2001}, respectively.

\begin{figure*}
\includegraphics{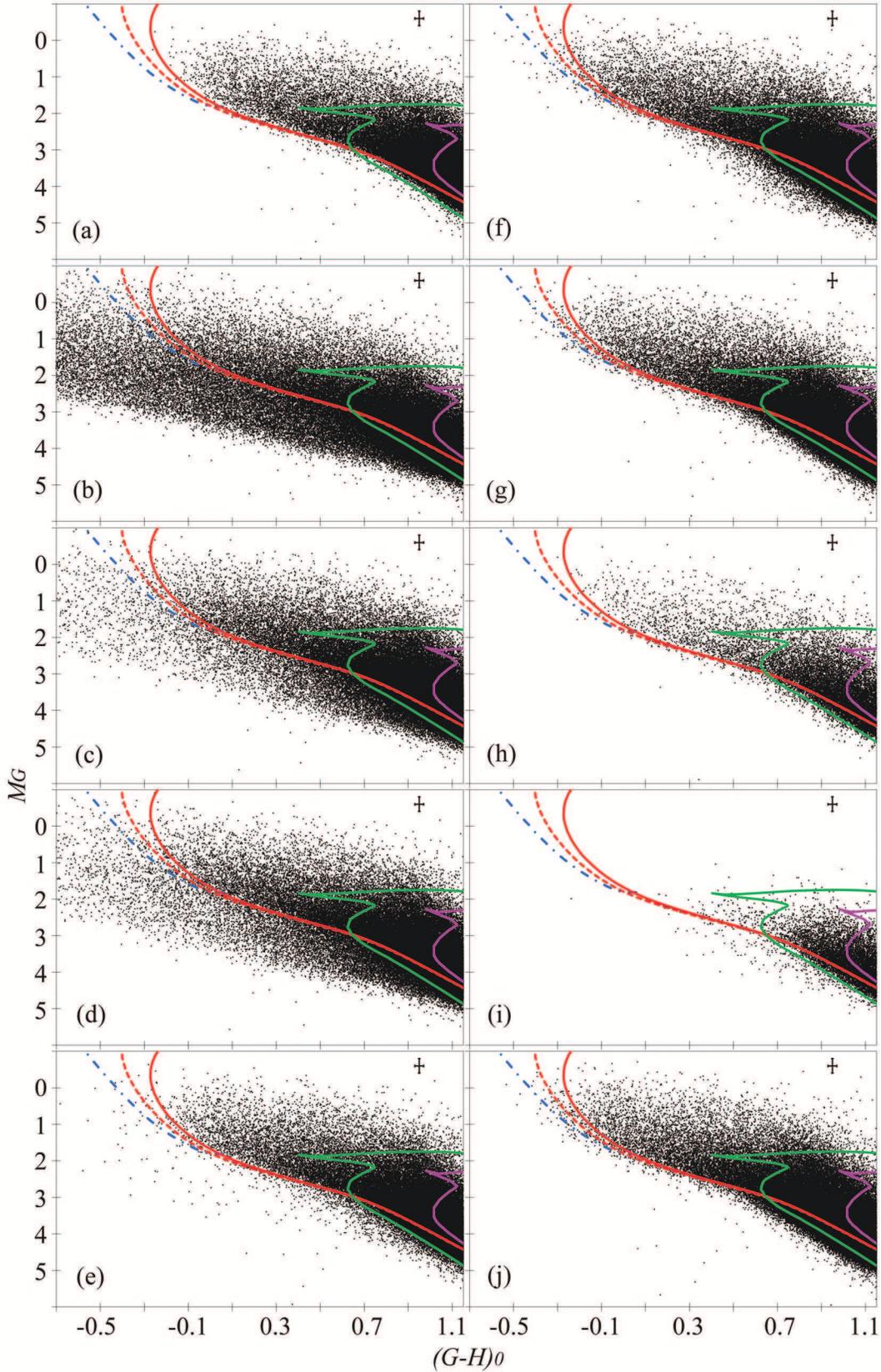}
\caption{HR diagram $(G-H)_0 - M_G$ for TGAS early-type main sequence stars with $\sigma(\varpi)/\varpi<0.1$.
The PARSEC isochrones are: 
3~Gyr, $\mathbf Z=0.0142$ -- purple line,
2.5~Gyr, $\mathbf Z=0.003$ -- green line,
200~Myr, $\mathbf Z=0.0152$ -- red solid line,
100~Myr, $\mathbf Z=0.0152$ -- red dotted line,
10~Myr, $\mathbf Z=0.0152$ -- blue dash-dotted line. 
$(G-H)_0$ and $M_G$ are corrected for reddening and extinction from the sources:
(a) no correction,
(b) PLA, 
(c) PLA$_R$,
(d) SFD$_R$,
(e) DCL,
(f) AGG,
(g) G12, 
(h) CLY,
(i) KKS,
(j) G17.
}
\label{gh}
\end{figure*}

\begin{figure*}
\includegraphics{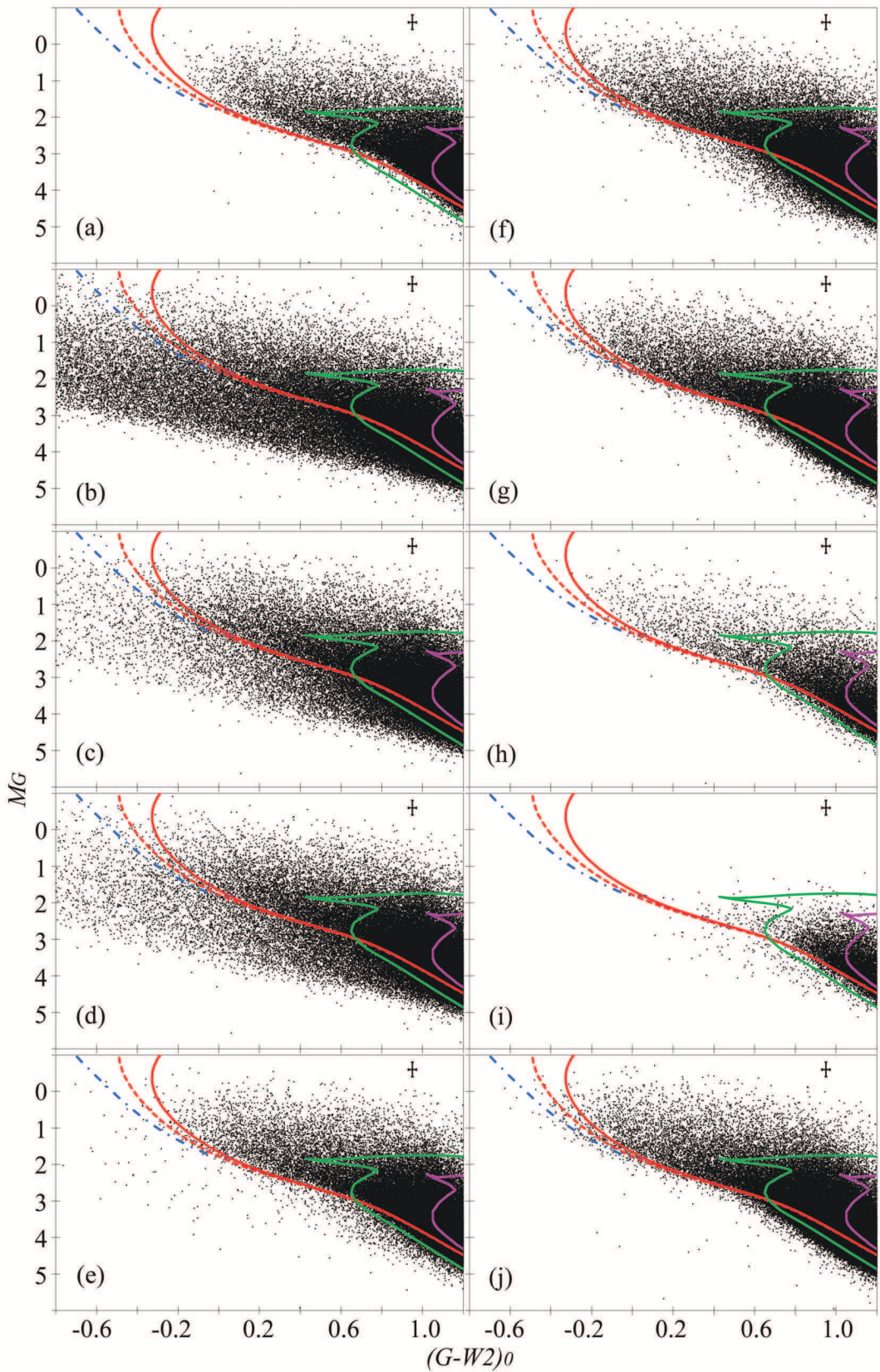}
\caption{The same as Fig.~\ref{gh} but for $(G-W2)_0$ and $M_G$.
}
\label{gw2}
\end{figure*}

\begin{figure*}
\includegraphics{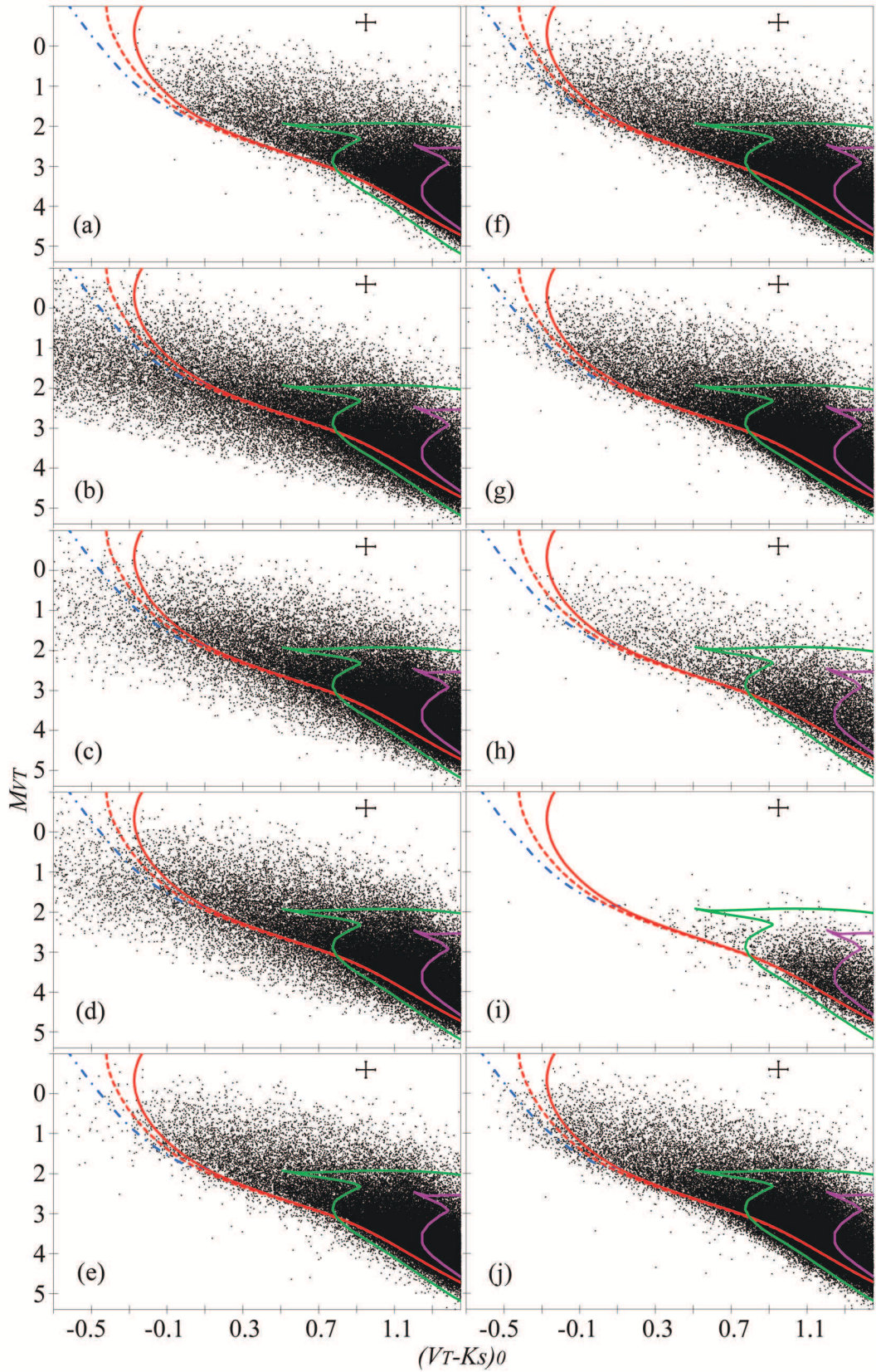}
\caption{The same as Fig.~\ref{gh} but for $(V_\mathrm{T}-K_s)_0$ and $M_\mathrm{V_T}$. 
}
\label{vtks}
\end{figure*}

\begin{figure*}
\includegraphics{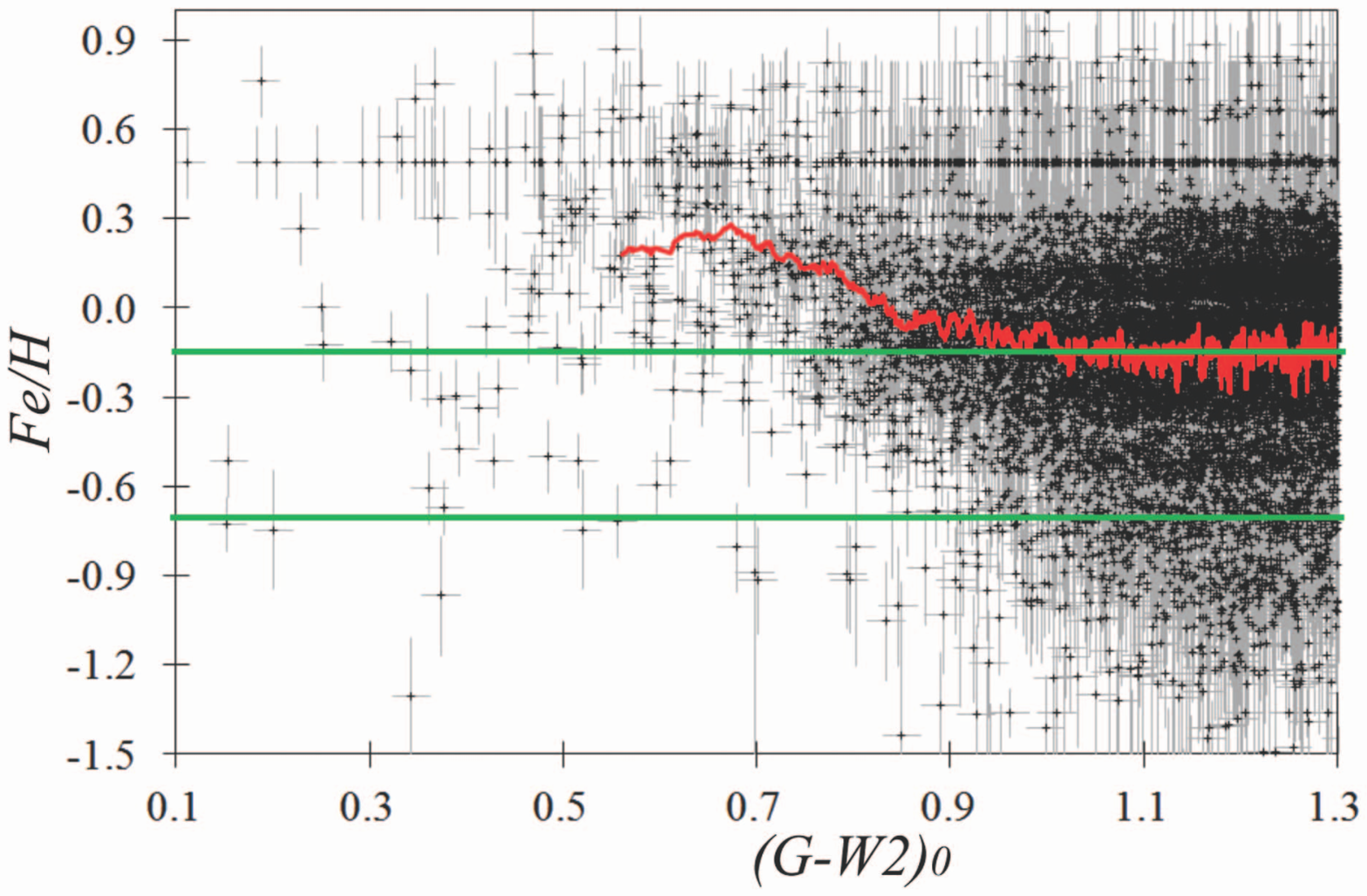}
\caption{$Fe/H$ vs. $(G-W2)_0$ with gray error bars for the TGAS stars with $\sigma(\varpi)/\varpi<0.1$ from the RAVE DR5 catalogue.
Red line is the moving average over 100 points.
The green lines show $Fe/H=-0.15$ and $-0.7$.
}
\label{gw2feh}
\end{figure*}

Obviously, the broad band photometry used cannot give us a precise age and metallicity for a star.
But the general distribution of stars with respect to the isochrones can be either right or wrong, in the sense whether it follows the BMG or other 
models of the Galaxy.

To maximize the influence of the reddening on color indices and, thus, to obtain more distinct positioning of the stars with respect to the isochrones
with less influence of the errors of parallax and photometry, we should use the widest color indices.

The distribution of the stars with respect to the isochrones can be different for different color indices due to 
(a) a deviation of the real extinction law from the accepted one, and
(b) some errors of the isochrones (an example of such an error for a PARSEC isochrone is given by \citet{g2017}).
Therefore, all possible color indices have been considered. They produce similar but slightly different results.
This signifies some deviation of the accepted extinction law and isochrones from reality.
However, this deviation is within the precision level of the data used and much lower than the differences between the estimates from the data sources
under consideration.
We conclude that the extinction law of \citet{cardelli} and both the PARSEC and MIST isochrones are acceptable at the current level of accuracy.

Three diagrams are presented here:
$(G-H)_0$ vs. $M_\mathrm{G}$ in Fig.~\ref{gh},
$(G-W2)_0$ vs. $M_\mathrm{G}$ in Fig.~\ref{gw2},
and
$(V_\mathrm{T}-K_s)_0$ vs. $M_\mathrm{V_T}$ in Fig.~\ref{vtks}.
The cut-off applied to $V_\mathrm{T}$ make the statistics and diagrams for $(G-H)_0$ and $(G-W2)_0$
more informative.
Every diagram shows the color indices and absolute magnitudes 
(a) uncorrected for the reddening and extinction as well as
corrected following the extinction law of \citet{cardelli} and the sources:
(b) PLA, 
(c) PLA$_R$,
(d) SFD$_R$,
(e) DCL,
(f) AGG,
(g) G12, 
(h) CLY,
(i) KKS,
(j) G17.
The diagrams with the SFD estimates are not shown because they are similar to the PLA diagrams.

The average errors of color and absolute magnitude due to those of photometry and parallax are shown in Fig.~\ref{gh}--\ref{vtks} by a separate dot with 
the error bars: 
\begin{equation}
\label{erro}
\left\{
\begin{aligned}
\sigma(G-H)=0.02,\\
\sigma(G-W2)=0.02,\\
\sigma(V_\mathrm{T}-K_s)=0.05,\\
\sigma(M_\mathrm{G})=0.17,\\
\sigma(M_\mathrm{V_T})=0.17.
\end{aligned}
\right.
\end{equation}

\begin{table*}
\def\baselinestretch{1}\normalsize\normalsize
\caption[]{The average and median $E(B-V)$ for the considered TGAS O--F stars younger than 3~Gyr and the declared errors $\sigma(E(B-V))$.
}
\label{ebverr}
\[
\begin{tabular}{lrrr}
\hline
\noalign{\smallskip}
Reddening from & $\overline{E(B-V)}$ & Median $E(B-V)$ & $\sigma(E(B-V))$ \\
\hline
\noalign{\smallskip}
SFD           & 0.38 & 0.10 & 0.03 \\
SFD$_R$ & 0.09 & 0.04 & 0.03 \\
PLA           & 0.39 & 0.11 & 0.03 \\
PLA$_R$ &  0.08 & 0.04 & 0.03 \\
AGG          &  0.07 & 0.05 & 0.03 \\ 
DCL           &  0.04 & 0.03 & 0.03 \\
G12           &  0.08 & 0.07 & 0.04 \\
CLY            &  0.05 & 0.04 & 0.03 \\	
KKS           &  0.04 & 0.02 & 0.03 \\	
G17           & 0.09 & 0.08 & 0.04  \\	
\hline
\end{tabular}
\]
\end{table*}


These bars do not reflect the errors of color and absolute magnitude due to those of reddening and extinction.
The errors $\sigma(E(B-V))$ declared by the authors of the corresponding data source are shown in Table~\ref{ebverr} together with 
the average and median $E(B-V)$ for all the considered TGAS O--F stars younger than 3~Gyr.
The latter is important because for many data sources the error in reddening depends on the reddening itself.
Usually, the authors declared the random but not the systematic errors.
The latter must be significant, as evident from the subsequent analysis of the distribution of the stars with respect to the isochrones in Fig.~\ref{gh},
\ref{gw2} and \ref{vtks}.
For the G17 map $\sigma(E(B-V))=0.04$~mag includes both the random and systematic errors as discussed in the Sect.~\ref{Gont_map}.
For the G12 model also the both kinds of errors are taken into account.
We note that for the SFD and SFD$_R$ maps the lower limit of $\sigma(E(B-V))=0.028$~mag is defined by the declared accuracy of the emission-to-reddening 
normalization.
A similar uncertainty defines the lower limit of the errors of the PLA and PLA$_R$ maps.
It is evident from the Table~\ref{ebverr} that the declared $\sigma(E(B-V))$ are of the same level of $0.03-0.04$~mag for all the data sources.
Thus, this is the current level of knowledge about the reddening within 280~pc from the Sun.

To select the isochrones for the comparison we have to estimate some average metallicity of the stars under consideration.
The majority of the O--F main sequence stars belong to the thin disc.
They exhibit various metallicities and some age--metallicity relation.
\citet{haywood} and \citet{marsakov} estimated the increase of the mean metallicity from about $0.012<\overline{\mathbf Z}<0.014$
\footnote{To avoid confusion, the metallicity is designated hereafter as $\mathbf Z$,
while one of the Galactic coordinates as $Z$.}
(i.e. $-0.04<\overline{Fe/H}<-0.10$ taking into account the solar $\mathbf Z=0.0152$ \citep{bressan})
to $\mathbf Z\approx0.015$ ($Fe/H=0$) for the last 3~Gyr.
The RAVE DR5 metallicity $Fe/H$ vs. de-reddened color $(G-W2)_0$ calculated with the extinction $A_\mathrm{V}$ 
from the RAVE DR5 is shown in Fig.~\ref{gw2feh} for the stars common to the TGAS and RAVE DR5 catalogues (some stars are outside the plot).
The grey bars show the precision of the data.
The red (grey) curve is the moving average over 100 points.
High mean metallicity for $(G-W2)_0<0.7$~mag may be a selection effect.
But, anyway, there are a considerable number of low metallicity stars at $(G-W2)_0>0.75$~mag.
The green (grey) horizontal lines show $Fe/H=-0.15$ ($\mathbf Z=0.0108$) and $Fe/H=-0.7$ ($\mathbf Z=0.003$).
The former is the mean $Fe/H$ for $(G-W2)_0>1$~mag.
The latter is a typical metallicity for the thick disc and, apparently,  the one for still a considerable number of stars in Fig.~\ref{gw2feh}.
Therefore, a noticeable number of stars can, and do, appear down to the isochrone 2.5~Gyr, 
$\mathbf Z=0.003$ being low metallicity subdwarfs.
This isochrone is shown in Fig.~\ref{gh}--\ref{vtks} by the green curve (the second from the right).
Such young low metallicity stars can be introduced to the solar vicinity by some radial migrations \citep{haywood} 
or merging of some satellites of the Galaxy \citep{sd}.

In agreement with all the estimates of metallicity, we accept the increase of the mean metallicity from $\mathbf Z\approx0.0142$ to $0.0152$
for the last 3~Gyr and show the key isochrones in Fig.~\ref{gh}--\ref{vtks}:
10~Myr, $\mathbf Z=0.0152$ -- by the blue dash-dotted curve (the first from the left),
100~Myr, $\mathbf Z=0.0152$ -- by the red dotted curve (the second from the left),
200~Myr, $\mathbf Z=0.0152$ -- by the red solid curve (the third from the left),
3~Gyr, $\mathbf Z=0.0142$ -- by the purple curve (the rightmost).
In the central and right parts of the figures, the 100 and 200~Myr isochrones are very close to each other and can be considered as the
solar metallicity zero-age main sequence (ZAMS).
The 10~Myr isochrone is only partly shown, namely to the left of the ZAMS. The rest part of this isochrone (which fits to the stars before their
setting on the ZAMS) corresponds to a negligible percentage of stars and is not shown.
We consider only those parts of the HR diagrams which approximately correspond to age less than 3~Gyr.

\begin{table*}
\def\baselinestretch{1}\normalsize\normalsize
\caption[]{The percentage of the outliers $2\sigma$ bluer than any reasonable PARSEC / MIST isochrone among all the stars younger than 3~Gyr.
}
\label{outliers}
\[
\begin{tabular}{lccc}
\hline
\noalign{\smallskip}
Reddening from & $(G-H)_0$ & $(G-W2)_0$ & $(V_\mathrm{T}-K_s)_0$ \\
\hline
\noalign{\smallskip}
No reddening & 0.05 / 0.10 & 0.10 / 0.15 & 0.08 / 0.11 \\
SFD           & 31 / 33 & 34 / 38 & 25 / 28 \\
SFD$_R$ & 11 / 12 & 13 / 15 &  6.9 / 8.1 \\
PLA           & 33 / 35 & 36 / 40 &  26 / 29  \\
PLA$_R$ & 8.1 / 8.9 & 11 / 13 &   5.0 / 5.9 \\
AGG          & 1.3 / 1.5 & 2.1 / 2.6 &  0.57 / 0.68 \\ 	
DCL           & 0.67 / 0.77 & 0.98 / 1.23 &  0.31 / 0.31 \\	
G12           & 0.34 / 0.41 & 0.54 / 0.68 &  0.34 / 0.40 \\	
CLY            & 0.26 / 0.28 & 0.29 / 0.34 & 0.28 / 0.34 \\	
KKS           & 1.2 / 1.4   & 1.8 / 2.4 &  0.71 / 0.98 \\	
G17           & 0.35 / 0.40 & 0.46 / 0.60 & 0.37 / 0.43 \\	
\hline
\end{tabular}
\]
\end{table*}


\begin{table*}
\def\baselinestretch{1}\normalsize\normalsize
\caption[]{The number of selected stars within $2\sigma$ from any reasonable PARSEC / MIST isochrone and younger than about 3~Gyr
for the data sources with full coverage of the sky.
}
\label{counts}
\[
\begin{tabular}{lrrr}
\hline
\noalign{\smallskip}
Reddening from & $(G-H)_0$ & $(G-W2)_0$ & $(V_\mathrm{T}-K_s)_0$ \\
\hline
\noalign{\smallskip}
No reddening & 35651 / 30561 &  24526 / 19737 & 29778 / 22778 \\
SFD           & 52145 / 48872 & 48227 / 42328 & 37608 / 32914 \\
SFD$_R$ & 49180 / 45467 & 43697 / 37257 &  38983 / 32657 \\
PLA           & 52401 / 49212 & 48276 / 42474 &  37491 / 32882 \\
PLA$_R$ & 51004 / 47204 & 45752 / 39022 &  40104 / 33690 \\
AGG          & 55595 / 51271 & 50988 / 43131 & 42832 / 36117 \\	
DCL           & 43723 / 39857 & 38263 / 31398 &  37686 / 30439 \\
G12           & 62412 / 57672 & 58877 / 49905 & 45973 / 39496 \\	
G17           & 65285 / 60493 & 62265 / 53178 &  46894 / 40605 \\	
\hline
\end{tabular}
\]
\end{table*}


It is evident from the Fig.~\ref{gh}--\ref{vtks} that some sources overestimate the reddening for many stars, producing some amount of outliers
which are bluer than any reasonable isochrone,
even outside the left sides of the plots for the PLA, PLA$_R$, SFD, and SFD$_R$ maps.
Taking into account the combined errors $\sigma$ of parallax and photometry, we calculate the percentage of the outliers located
more than $2\sigma$ to the left of the 10~Myr, $\mathbf Z=0.0152$ or 100~Myr, $\mathbf Z=0.0152$ or 2.5~Gyr,
$\mathbf Z=0.003$ isochrones among all the stars younger than 3~Gyr.
The results for all the data sources, both the PARSEC and MIST isochrones and three color indices are presented in Table~\ref{outliers}.
One can see that the DCL, G12, CLY, G17 and zero reddening case have the lowest percentage of outliers, 
whereas the remaining sources provide slightly more (AGG and KKS) or much more (SFD, SFD$_R$, PLA, PLA$_R$) outliers.
All the color indices and both the PARSEC and MIST isochrones give similar results.
It is evident that the estimates of the reddening to infinity taken from the PLA and SFD maps are far from precise.
Even reduced to the distances by use of the barometric law of the dust distribution, the estimates from the PLA$_R$ and SFD$_R$
maps are still unacceptable.
Consequently, estimating reddening inside the dust layer using estimates from outside it is a major issue. 

It should be emphasized that the outliers are not the only stars with wrong estimates of reddening and extinction.
They are those which are easily revealed in the HR diagrams.
Many more stars with wrong reddening and extinction estimates might hide in the bulk of the stars in the diagrams.

Firstly, in all the figures the plots without any correction for reddening and extinction show few (if any) stars at 10, 100 and 200~Myr isochrones
in the left and middle parts of the plots and even some narrow gap between the bulk of the stars and the 200~Myr isochrone.
This means that there are too few stars younger than 200~Myr with respect to the BMG or any other model of the Galaxy.
This proves that the accuracy of the parallax and photometry is high enough to show that the color indices and absolute magnitudes should be corrected
for reddening and extinction.

Secondly, the same but smaller deficit of the youngest stars is seen for the DCL and CLY maps in the plots (e) and (h), respectively.
Apparently, in these cases the reddening is underestimated for many stars located between the 10~Myr and 3~Gyr isochrones.
We can count such stars to test this underestimation.
In Table~\ref{counts} for three color indices and data sources with full coverage of the sky (i.e. except CLY and KKS),
we list the number of all the stars younger than 3~Gyr minus the number of the outliers which were counted earlier.
Taking into account all possible uncertainties, BMG predicts that 20--30~per cent (48,300--72,500 stars) of the TGAS stars with
$\sigma(\varpi)/\varpi<0.1$ (241,681 stars) have ages less than 3 Gyr and should be taken into account in Table~\ref{counts}.
The rest are mainly giants and G dwarfs.
The counts for $(V_\mathrm{T}-K_s)_0$ may be underestimated due to larger errors and the magnitude cut.
However, the counts for $(G-H)_0$ and $(G-W2)_0$ show that only the AGG, G12 and G17 provide a reliable number of  young stars
(formally, the PLA and SFD maps also provide similar results, but they have been criticized because of the large number of outliers).
The CLY map showing the least percentage of the outliers might also fit the BMG's percentage of the O--F stars, but
since it covers too little area of the sky, we cannot make our final conclusion on this map.

As noticed earlier, the KKS (RAVE DR5) estimates are the only direct measurements of the extinction for the TGAS stars.
Comparing the KKS results with others we can conclude that at the current level of accuracy there are no advantages of the direct
measurements with respect to the 3D maps and models, i.e. estimates for stars close to the TGAS ones.

\section{Conclusions}
\label{conclusions}

We studied the location of approximately 60,000 {\it Gaia} DR1 TGAS O--F main sequence stars younger than 3~Gyr in the HR diagrams
in comparison with the PARSEC and MIST theoretical isochrones.
We used the distances from \citet{bailer3} for the TGAS stars with the most precise parallaxes ($\sigma(\varpi)/\varpi<0.1$) together with
the precise (at the level of $0.02$~mag) photometry from the {\it Gaia} $G$ band, {\it Tycho-2} $V_\mathrm{T}$ band, 2MASS $H$, $K_s$ bands, and 
{\it WISE} $W2$ band.
To calculate the color indices and absolute magnitudes, we took into account 2D and 3D reddening and extinction estimates from eight sources.

These sources have different angular and spatial resolution.
The DCL and G17 have quite limited angular resolution, whereas the angular resolution of the PLA and SFD maps is very high.
The AGG and G12 are, in fact, analytical models which describe the reddening and extinction by formulas.
Therefore, formally they provide an estimate for any star. In practice, however, they merely interpolate in a grid which may have low resolution.
For example, using the AGG model, one can extract the same extinction values for stars which are located in wide regions at the Galactic polar caps.
This means that the real resolution of this model is several degrees and tens of parsecs.
None the less, even such large differences in angular and spatial resolutions have little impact on the results.
Moreover, for the KKS the resolution is not important at all, due to the direct measurement of extinction for the selected TGAS stars
and the usage of their TGAS parallaxes instead of the estimates of their distances from RAVE DR5.
However, the KKS estimates certainly are not the best ones according to our results. This proves that at the current level of accuracy the resolution 
is not very important.

The reddening estimates under consideration have different origins. The SFD and PLA maps are based on the measurements of the IR emission from the observer to infinity 
which is converted to $E(B-V)$ by correlating with color excesses of galaxies, quasars, and stars. 
Up till now these maps represent the most precise approach to estimating the reddening and extinction for objects outside the dust layer of our Galaxy.
Yet, to use them inside the layer (i.e. to convert them from 2D into 3D)
one should at least take into account the dust distribution and spatial variations of the extinction law in the layer.
The simplest description of the dust distribution is the barometric law which has been applied in this study to convert the SFD and PLA maps to the 3D ones.
A more sophisticated attempt to calculate the reddening and extinction inside the dust layer based on the SFD or any other emission-based map is
the DCL map.
The KKS used the SFD map as a prior, whereas the CLY used it to derive the reference sample of stars.
The remaining 3D sources, namely, AGG, G12 and G17 are independent of any emission-based map and based solely on the multi-color photometry.
It is evident from the HR diagrams that only these independent data sources show reliable estimates of reddening and extinction. 
This may due to a complicated distribution of the dust. 
An estimate of reddening and extinction outside the dust layer is not enough to estimate these quantities inside of the layer.
The SFD and PLA maps, even reduced with the barometric law for the dust distribution, produce incorrect estimates inside the layer.
SFD-based 3D map of \citet{green} is not applicable within a few hundred pc from the Sun.
Another SFD-based, the DCL map underestimates the reddening for many stars, producing too few O--F and too many G dwarfs.
Also, the failure of the DCL map, despite of its adjustment to the local space, shows that the astronomical community should pay more attention 
to the description of the local dust medium. 
The authors of the DCL noted: `Ideally this region about the Sun should be described by a more detailed local model of the dust distribution.'
The physically grounded model of G12, which describes the Gould Belt as a local container of dust (see Sect.~\ref{Gont_model}) deserves more attention and 
further development.
We conclude that the AGG and G12 models and G17 map with detailed description of some local variations of the dust medium are better than other
data sources which we have considered in this study
for estimating the reddening and extinction of the TGAS stars embedded in the dust layer within 280 pc from the Sun.
However, the AGG, G12 and G17 may fail in describing the reddening and extinction in distant regions of the Galactic disc.
Probably, the other sources would provide better estimates there.

For the majority of the TGAS stars the data sources under consideration consistently show some considerable differences between the
reddening/extinction estimates to the stars and to infinity along the same lines of sight.
This means that the majority of the TGAS stars are located inside the dust layer.
A significant proportion of all {\it Gaia} stars may appear inside the dust layer.
Thus, this study may be relevant not only for the {\it Gaia} DR1 but also for future {\it Gaia} releases.

It is worth noting, that the AGG model from 1992 still describes the dust medium better than some later and more sophisticated models and maps.
If the accuracy of the extinction remains at the same level, the usage of the final {\it Gaia} data release will be limited by the uncertainty 
of the reddening and extinction:
the expected uncertainty of parallax of 1~per cent translates into those of absolute magnitude of 0.02~mag, whereas due to the 
extinction this uncertainty may be 0.1~mag, on average.

\section{Acknowledgements}

We thank an anonymous reviewer for useful comments.
We thank Justin Turman, Leslie Morrison and Leonid Petrov for the improvement of the style of this paper.
AVM is partly supported by the Russian Foundation for Basic Researches (grant number 14-02-810).
AVM is a beneficiary of a mobility grant from the Belgian Federal Science Policy Office.
The resources of the Centre de Donn\'ees astronomiques de Strasbourg, Strasbourg, France 
(\url{http://cds.u-strasbg.fr}) including the reddening/extinction data sources
under consideration were widely used in this study.
This work has made use of data from the European Space Agency (ESA) mission {\it Hipparcos}.
This publication makes use of data products from the Two Micron All Sky Survey, which is a joint project of the University of Massachusetts and 
the Infrared Processing and Analysis Center/California Institute of Technology, funded by the National Aeronautics and Space Administration and
the National Science Foundation.
This publication makes use of data products from the Wide-field Infrared Survey Explorer, which is a joint project of the University of California, 
Los Angeles, and the Jet Propulsion Laboratory/California Institute of Technology, funded by the National Aeronautics and Space Administration.
This work makes use of data products from the Radial Velocity Experiment (RAVE).
Funding for RAVE (\url{www.rave-survey.org}) has been provided by institutions of the RAVE participants and by their national funding agencies.
This work has made use of data from the European Space Agency (ESA) mission {\it Gaia} (\url{https://www.cosmos.esa.int/gaia}), processed by
the {\it Gaia} Data Processing and Analysis Consortium (DPAC, \url{https://www.cosmos.esa.int/web/gaia/dpac/consortium}). Funding
for the DPAC has been provided by national institutions, in particular the institutions participating in the {\it Gaia} Multilateral Agreement.

\bibliographystyle{mn2e}
\bibliography{art7_arxiv}

\label{lastpage}
\end{document}